\documentclass[aps,prl,preprint,superscriptaddress]{revtex4-1}
\usepackage[utf8]{inputenc}
\usepackage{graphicx}
\usepackage{amsmath}

\begin{document}


\title{Optical control of vibrational coherence triggered by an ultrafast phase transition}


\author{M. J. Neugebauer}
\email{mj.neugebauer@phys.ethz.ch}
\author{T. Huber}
\author{M. Savoini}
\author{E. Abreu}
\affiliation{Institute for Quantum Electronics, Physics Department, ETH Zurich, CH-8093 Zurich, Switzerland}
\author{V. Esposito}
\affiliation{Swiss Light Source, Paul Scherrer Institut, CH-5232 Villigen PSI, Switzerland}
\author{M. Kubli}
\affiliation{Institute for Quantum Electronics, Physics Department, ETH Zurich, CH-8093 Zurich, Switzerland}
\author{L. Rettig}
\altaffiliation{Current address: Abteilung Physikalische Chemie, Fritz-Haber-Institut der Max-Planck-Gesellschaft, D-14195 Berlin, Germany}
\author{E. Bothschafter}
\author{S. Grübel}
\affiliation{Swiss Light Source, Paul Scherrer Institut, CH-5232 Villigen PSI, Switzerland}
\author{T. Kubacka}
\affiliation{Institute for Quantum Electronics, Physics Department, ETH Zurich, CH-8093 Zurich, Switzerland}
\author{J. Rittmann}
\author{G. Ingold}
\author{P. Beaud}
\affiliation{Swiss Light Source, Paul Scherrer Institut, CH-5232 Villigen PSI, Switzerland}
\author{D. Dominko}
\affiliation{Institute for Physics, Johannes Gutenberg Universität Mainz, D-55128 Mainz, Germany}
\affiliation{Institute of Physics, HR-10000 Zagreb, Croatia}
\author{J. Demsar}
\affiliation{Institute for Physics, Johannes Gutenberg Universität Mainz, D-55128 Mainz, Germany}
\author{S. L. Johnson}
\affiliation{Institute for Quantum Electronics, Physics Department, ETH Zurich, CH-8093 Zurich, Switzerland}


\date{\today}

\begin{abstract}
Femtosecond time-resolved x-ray diffraction is employed to study the dynamics of the periodic lattice distortion (PLD) associated with the charge-density-wave (CDW) in K\(_{0.3}\)MoO\(_3\). Using a multi-pulse scheme we show the ability to extend the lifetime of coherent oscillations of the PLD about the undistorted structure through re-excitation of the electronic states. This suggests that it is possible to enter a regime where the symmetry of the potential energy landscape corresponds to the high symmetry phase but the scattering pathways that lead to the damping of coherent dynamics are still controllable by altering the electronic state population. The demonstrated control over the coherence time offers new routes for manipulation of coherent lattice states.  
\end{abstract}

\pacs{}

\maketitle


The use of ultrashort laser pulses to generate and manipulate coherent states of lattice vibrations has been demonstrated in a wide variety of crystalline materials~\cite{Merlin1997,Johnson2017}. Typically, the largest responses are obtained when the pulse photon energy is tuned to a region of pronounced absorption in the material, triggering electronic transitions that strongly couple to small wavevector vibrational modes. This is often referred to as ``displacive excitation of coherent phonons'' (DECP), in the limit where the light absorption happens on timescales shorter than the period of resulting vibrations \cite{Cheng1991,Zeiger1992a,Garrett1996}.  The DECP mechanism is often understood in terms of a time-dependent interatomic potential energy surface for the crystal ions. The fast absorption induces a sudden shift in the quasiequilibrium structure of the crystal which excites a coherent oscillation of a normal mode about a displaced coordinate.  Several experiments have demonstrated coherent control of these oscillations in different materials using a multi-pulse scheme to further shift the quasiequilbrium structure at controlled time delays~\cite{Hase1996,Roeser2004,DeCamp2001,Beaud2007,Rettig2014}, under low-fluence conditions where the displacement is approximately proportional to the excitation fluence. 

In some situations strong optical excitation can lead to changes in the overall symmetry of the interatomic potential, a phenomenon that is often identified as an ``ultrafast’’ phase transtion~\cite{Beaud2009,Eichberger2010,Lu2010,Huber2014,Beaud2014,Trigo2018}. In some cases the symmetry change is short-lived and collapses back into the low-symmetry structure within a few picoseconds~\cite{Yusupov2010}. In this situation multiple pulse excitation enables the study of the dynamically evolving potential surface by inducing DECP in the partially relaxed structure~\cite{Wall2012}. In other cases, under strong enough excitation conditions and/or long-lived electronic and structural excitations, the change in symmetry persists up to microseconds. Typically, the system then relaxes back to the low-symmetry state only after thermalization and heat transport have led to cooling the material back to its initial temperature. Several experiments have studied this regime and observed dynamics in the high-symmetry structural configuration~\cite{Huber2014, Beaud2014, Trigo2018}. Beyond that the possibility of controlling coherent oscillations within the high-symmetry phase remains largely unexplored. Here we focus on this issue, exploring possible avenues of control over the dynamics that follow the light-driven collapse of the CDW order in K\(_{0.3}\)MoO\(_3\), a model system for a one-dimensional  Peierls transition \cite{Peierls1955}.

In equilibrium, K\(_{0.3}\)MoO\(_3\) undergoes a metal-to-insulator transition at  \(T\textsubscript{c} = 183\)~K, accompanied by the formation of a CDW~\cite{Fogle1972,Travaglini1981}. This transition is preceeded by a Kohn anomaly~\cite{Pouget1991}. Strong excitation with a femtosecond optical pulse can melt the CDW, inducing a phase transition to the metallic state. Experiments using optical reflectivity as a probe show either a disappearance of amplitude mode oscillations~\cite{Tomeljak2009} or a dramatic softening and increase in damping~\cite{Mankowsky2017} above a critical absorbed fluence of \(F\textsubscript{c}\textsuperscript{opt} \approx 0.3\)~mJ/cm\(^2\) for pump pulses at a wavelength \(\lambda = 800\)~nm. Experiments using x-rays to probe directly the collapse of the PLD estimate a critical fluence of \(F\textsubscript{c}\textsuperscript{x-ray}\approx 1.0\)~mJ/cm\(^2\) \cite{Huber2014}, which is roughly comparable to \(F\textsubscript{c}\textsuperscript{opt}\), especially considering differences in the probing methods. For excitation fluences \(F\geq 1.5\cdot F\textsubscript{c}\textsuperscript{x-ray}\) the PLD does not simply vanish but transiently revives after around 0.3~ps, which is ascribed to coherent dynamics along the Peierls coordinate~\cite{Huber2014}. These dynamics correspond to a pair of acoustic modes with the wavevector of the Peierls distortion but in a quasi-equilibrium structure with symmetry equivalent to the metallic phase. The coherent dynamics exhibit an unusual damping behavior, resulting in an abrupt stop of coherent motion after only half a vibrational period. This appears to be inconsistent with the normal assumption of viscous damping that typically results from perturbative coupling to other excitations~\cite{Huber2014}. 

These observations open the question of whether some degree of control of these coherent dynamics in the high-symmetry phase is possible, despite the fact that the long wavevector of the underlying acoustic modes normally precludes further displacive optical excitation. We explore this question using a two-pulse excitation scheme: While the first pump melts the electronic order and launches the coherent motion, the second re-excites the system during the motion. We study with time-resolved x-ray diffraction how the re-excitation of the second pulse affects the coherent dynamics.

For our experiments we use a bulk sample of K\(_{0.3}\)MoO\(_3\) cleaved along its \((2~0~\overline{1})\) plane and cooled with a nitrogen blower to 95~K, substantially below \(T\textsubscript{c}\). The PLD associated with the CDW can be probed using hard x-ray diffraction by monitoring the intensity of the \((1~(4-\nolinebreak[4]q\textsubscript{b})~\overline{0.5})\) superlattice Bragg reflection, where \(q\textsubscript{b}\) is the modulation wavevector along the chain direction (\(b\)-axis). At 100~K the modulation wavevector is \(q\textsubscript{b} = 0.748(1)\) \cite{Schutte1993}. In the kinematic approximation the diffraction intensity is proportional to the square of the magnitude of the PLD.

\begin{figure}
\includegraphics[width= 10 cm]{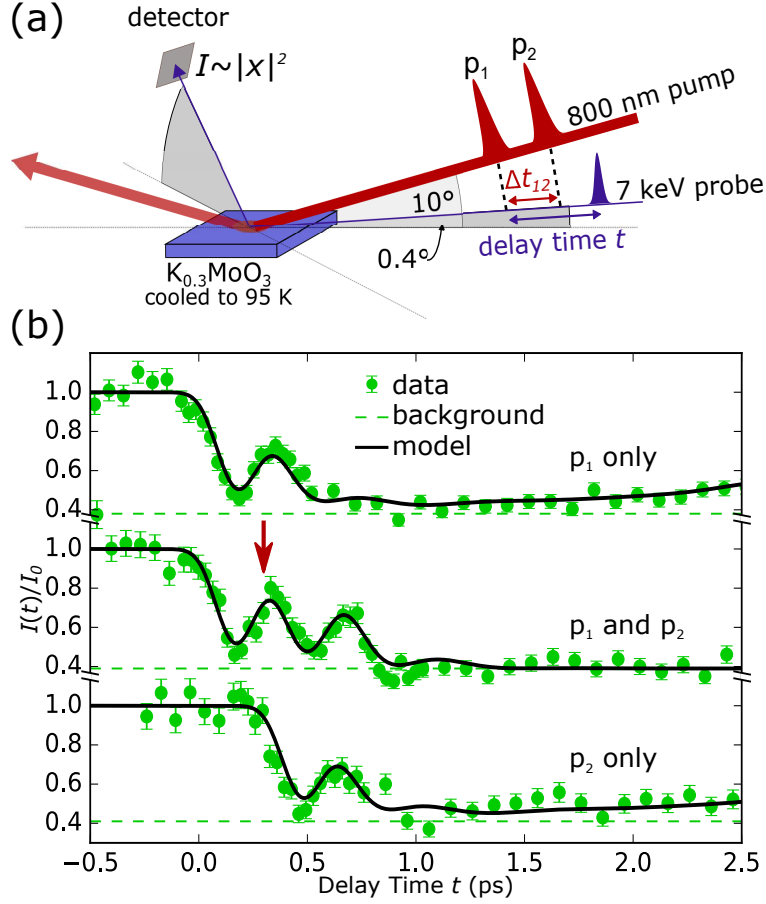}
\caption{\label{fig:double_pump} Scheme of the experimental setup in (a) and comparison between individual and sequential application of  the pump pulses in (b). The two different delay times are indicated in (a). One is the delay time \(t\) between the first pump pulse \(p_1\) and the x-ray probe, while \(\Delta t\textsubscript{12}\) is the time delay between the two pump pulses. In (b) the trace with the double-pulse excitation (middle) is compared to the traces obtained by excitations with individual \(p_1\) (top) and \(p_2\) (bottom) pulses at \(F_1 = F_2 = F_0\). Dashed lines indicate the background level, black lines correspond to the model (cf. text).}
\end{figure}

A sketch of the experimental setup is presented in Fig. \ref{fig:double_pump}(a). The structural dynamics associated with the CDW-state are investigated using 7 keV x-ray pulses with a FWHM-duration of around 120~fs and the sample is excited with 100~fs~(FWHM) \(p\)-polarized 800~nm laser pulses. A Mach-Zehnder scheme creates a second pump pulse \(p_2\), which can be delayed by \(\Delta t\textsubscript{12}\) relative to the first pump pulse \(p_1\). In order to match the penetration depths of the optical  and x-ray  beams a grazing incidence geometry is chosen. We set \(F_1\) to \(F_0= 1.7\)~mJ/cm\(^2\) to be above the critical fluence of the previous study \cite{Huber2014}, while \(F_2\) varies between \(F_0/4\) and \(F_0\). We estimate the experimental time resolution to be 150~fs (see Supplementary Information). 

Fig. \ref{fig:double_pump}(b) shows the  time evolution of the superlattice diffraction intensity for excitation with each pulse individually as well as both sequentially. If only \(p_1\) or \(p_2\) are applied at a fluence of \(F_0 = 1.7\)~mJ/cm\(^2\), a single transient revival appears around 0.30~ps after the arrival of the excitation pulse, in agreement with the results of Ref. \cite{Huber2014} (cf. Fig. \ref{fig:F_scan}). The middle plot shows the time evolution when both \(p_1\) and \(p_2\) are present and \(\Delta t\textsubscript{12}=0.30\)~ps (the arrival of \(p_2\) is indicated with red arrows in all plots). Here, a second revival of the CDW-distortion is visible at \(t\approx 0.60\)~ps, whose shape and magnitude resemble the first one. A background level intensity \(I\textsubscript{bg}\) remains in the superlattice diffraction peak even for high excitation fluence. We ascribe this to the fraction of unexcited volume of the sample that is probed by the x-rays, as already reported in Ref. \cite{Huber2014}. In all plots the background level \(I\textsubscript{bg}\) fit to the model curves is shown as a dashed line. 

\begin{figure}
\includegraphics[width= 10 cm]{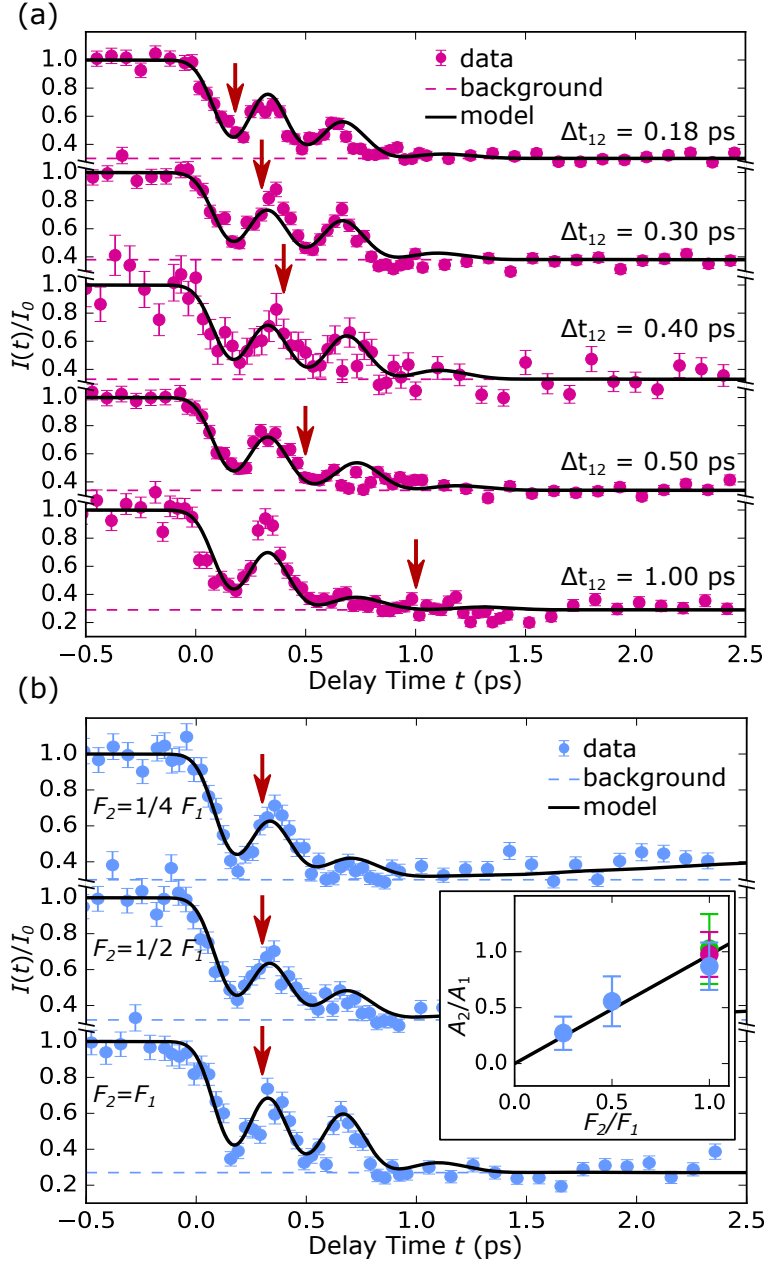}
\caption{\label{fig:p_2_var} The data taken in the double-pulse excitation configuration. (a) The dependence on \(\Delta t\textsubscript{12}\) for fixed \(F_1 = F_2 = F_0\). (b) The dependence on \(F_2\) for fixed \(\Delta t\textsubscript{12}=0.30\)~ps. Dashed lines indicate the background level, black lines correspond to the model (cf. text). The inset of (b) shows the dependence of the ratio between the amplitude of the second (\(A_2\)) and the first (\(A_1\)) revival on the relative fluence of the second excitation for \(\Delta t\textsubscript{12}=0.30\)~ps including a linear fit. The colors of the data points indicate the corresponding time trace (see also Fig. \ref{fig:double_pump}).}
\end{figure}

We now focus on the temporal evolution of the PLD as a function of the re-excitation delay \(\Delta t\textsubscript{12}\) between 0.18~ps and 1.00~ps with \(F_2 = F_1\), as shown in Fig. \ref{fig:p_2_var}(a). Clearly, the magnitude of the second revival depends on \(\Delta t\textsubscript{12}\), with a maximum near \(\Delta t\textsubscript{12} = 0.30~\)ps. A further increase of \(\Delta t\textsubscript{12}\), e.g. to \(\Delta t\textsubscript{12} = 0.50~\)ps or 1.00~ps, leads to no clear additional response of the system. Furthermore, \(F_2\) is also varied while keeping \(\Delta t\textsubscript{12}\) at 0.30~ps. The resulting delay time scans for \(F_2 = F_1, F_1/2\), and \(F_1/4\) are displayed in Fig. \ref{fig:p_2_var}(b). We define the amplitude of the first revival \(A_1\) as the difference between its maximum and the minimum of the first half-cycle, and the second revival amplitude \(A_2\) accordingly. The ratio of \(A_2/A_1\) scales linearly with \(F_2\), as shown in the inset. Additionally, we show \(A_2/A_1\) for \(\Delta t\textsubscript{12} = 0.30~\)ps and \(F_2 = F_1\) from the other two data sets (cf. Fig. \ref{fig:double_pump} and \ref{fig:p_2_var}(a), colors correspond) to underline the similar amplitude of the two revivals for this configuration. The timing of the revivals are, within our experimental uncertainties, independent of \(\Delta t\textsubscript{12}\) and \(F_2\).

To describe the dynamics we extend the phenomenological model of Ref.~\cite{Huber2014}. The concept is similar to that of the Landau theory for second order phase transitions \cite{Landau1968}, where we define a phenomenological parameterization of an effective ionic potential energy surface rather than a free energy.  The basic idea is that the shape of the effective potential depends strongly on the  electronic states that are populated at a given time after the optical excitation.  For simplicity we will consider a potential 
\begin{eqnarray}
V(x) = \frac{1}{2} a x^2 + \frac{1}{4} b x^4
\end{eqnarray}
where \(a\) and \(b\) are parameters, and \(x\) is a structural coordinate giving the instantaneous magnitude of the PLD associated with the CDW.  As in Ref.~\cite{Huber2014},  we consider the parameter \(a\) to be a function of the electronic state of the material and the parameter \(b\) to be constant. For convenience we will work in dimensionless units for \(V\) and \(x\) where \(b = 1\) and \(a = -1\) for the ground state of the material. For these choices, the minima of \(V(x)\) in the ground state occur at \(x\textsubscript{min} = \pm 1\). Without loss of generality we will assume that the equilibrium ground state value is \(x_0=1\). For a more general value of \(a\) we have either \(x\textsubscript{min} = \pm \sqrt{a}\) for \(a < 0\) or \(x\textsubscript{min} = 0\) for \(a \geq 0\). We can identify \(x\textsubscript{min}\) as an effective order parameter of the CDW phase. 

The electronic excitation of the material from the laser interaction will cause \(a\) to become time-dependent. In Ref.~\cite{Huber2014} \(a\) was assumed to depend linearly on a dimensionless  electronic energy density parameter \(\eta\) that depends on the excitation fluence. While this may be appropriate for low or moderate excitation levels, at high excitation levels we encounter a problem since allowing an arbitrarily large value of \(a\) gives unrealistically high frequencies for vibrations along the PLD coordinate \(x\) for strong excitation levels. We will therefore make a rough approximation for \(a(\eta)\) that  prevents this effect by defining
\begin{eqnarray}
a(\eta) = 
\begin{cases} 
\eta - 1 & \text{if }\eta < 1 + a\textsubscript{max}\\
a\textsubscript{max} & \text{if }\eta \geq 1 + a\textsubscript{max}
\end{cases}
\end{eqnarray}
where \(a\textsubscript{max} > 0\) is a constant.  

The excitation parameter \(\eta\) depends on time, depth \(z\) from the sample surface, and the strength of the pump pulse(s). For a single excitation pulse at \(t=0\), we approximate \(\eta\) as
\begin{eqnarray}
\eta\textsubscript{S}(z,t) = \eta_0 e^{-z/\delta_L} e^{-t/\tau\textsubscript{disp}}\Theta(t)
\end{eqnarray}
where \(\eta_0\) is a dimensionless parameter depending on the pump fluence \(F_1\). \(\delta\textsubscript{L}\) is the \(1/e\) penetration depth of the laser intensity, \(\tau\textsubscript{disp}\) is a relaxation time, and \(\Theta\) is the Heaviside step function. If we now add a second pulse with fluence \(F_2\) separated by a time \(\Delta t_{12}\), we have instead
\begin{widetext}
\begin{eqnarray}
\eta\textsubscript{D}(z,t) = \eta_0 e^{-z/\delta_L} \left[ \Theta(t) e^{-t/\tau\textsubscript{disp}} + \Theta(t-\Delta t_{12})\frac{F_2}{F_1} e^{-(t-\Delta t_{12})/\tau\textsubscript{disp}}\right].
\end{eqnarray}
\end{widetext}
The duration of the excitation pulses is taken into account by a convolution with Gaussian of 0.10~ps FWHM.

The equation of motion for \(x\) is
\begin{eqnarray}
\ddot{x} = -\omega^2 \left[ a(t) x + x^3 \right] - 2 \gamma(t) \dot{x}
\label{eq:eom}
\end{eqnarray}
where \(\omega = 2 \pi \nu\), \(\nu = 1.53\)~THz is the amplitude mode frequency in the ground state~\cite{Huber2014} and \(\gamma(t)\) is a phenomenological damping coefficient.  As discussed in Ref.~\cite{Huber2014}, in order to make it possible to fit Eq.~\ref{eq:eom} to the dynamics we observe experimentally, \(\gamma\) should be suppressed for a short time after the pulse.  Microscopically, this would correspond to fewer scattering channels from the amplitude mode available under conditions of very high electronic excitation. Using arguments analogous to our form for \(a(\eta)\), we consider this transient suppression of damping to be of the form
\begin{eqnarray}
\gamma(z,t) = 
\begin{cases}
\gamma^*(z,t) & \text{if }\gamma^*(z,t) > \gamma\textsubscript{min}\\
\gamma\textsubscript{min} & \text{otherwise}
\end{cases}
\end{eqnarray}
with
\begin{widetext}
\begin{eqnarray}
\gamma^*(z,t)=\gamma\textsubscript{unex}\Theta(-t) + \gamma_0\left[1-g (a+1) e^{-z/\delta_L}\left(\Theta(t)e^{-t/\tau_\gamma}+\frac{F_2}{F_1}\Theta\left(t-\Delta t\textsubscript{12}\right)e^{-\left( t-\Delta t\textsubscript{12}\right)/\tau_\gamma}\right)\right],
\label{eq:gamma_star}
\end{eqnarray}
\end{widetext}
where \(g\) is a dimensionless constant and \(\tau_\gamma\) is a relaxation time scale. The constants \(\gamma\textsubscript{unex}\) and \(\gamma\textsubscript{min}\) are introduced as the damping value before excitation and the minimum permissible value for the transient damping parameter respectively. The latter prevents the damping from becoming unreasonably small (or even negative) at high excitation values. Physically, \(\gamma\textsubscript{min}\) represents alternative scattering channels that are not suppressed by the electronic excitation. We set \(\gamma\textsubscript{unex}\) to 0.4~ps\(^{-1}\) and \(\gamma\textsubscript{min}\) to 0.2~ps\(^{-1}\) - see Supplementary Information. We can now solve Eq.~\ref{eq:eom} with initial conditions \(x = x_0\) and \(\dot{x} = 0\) to find \(x\) as a function of both time \(t\) and depth \(z\). 

The intensity of x-ray diffraction from the superlattice peak is proportional to a weighed average of \(x(z,t)\) over the 1/e attenuation length \(\delta_X = 100 \text{ nm}\) of the x-ray intensity
\begin{eqnarray}
\frac{I(t)}{I_0} =  \frac{1}{\delta_X} \int_0^\infty x^2(z,t) e^{-2z/\delta_X} dz
\end{eqnarray}
which we then convolve with a Gaussian of FWHM 150~fs to approximate the experimental time resolution. The result we compare directly with the data.

\begin{figure}
\includegraphics[width= \textwidth]{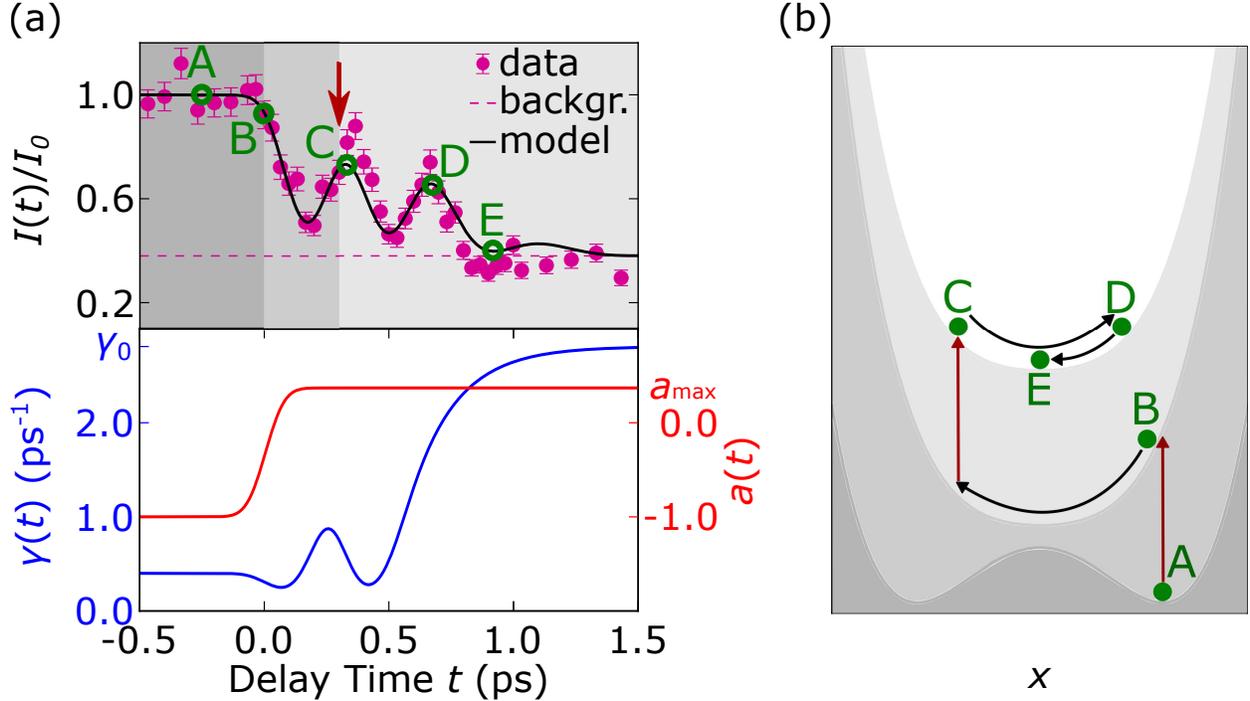}
\caption{\label{fig:expl_plot} Visualization of the model described in the text. The upper curve of (a) is a typical pump-probe trace with \(\Delta t\textsubscript{12}=300\)~fs and \(F_2 = F_1\) (Dashed lines indicate the background level, black lines correspond to the model (cf. text)). Below the corresponding progression of the damping \(\gamma\) (blue, left axis) and \(a(t)\) (red, right axis) near the surface are shown. (b) Sketch of the evolution of the PLD in its potential at delay times denoted on the pump-probe trace. The background colors indicate the potential configuration the system is in at different delay times. The two high symmetry potentials are offset for clarity.}
\end{figure}

\begin{figure}
\includegraphics[width=12cm]{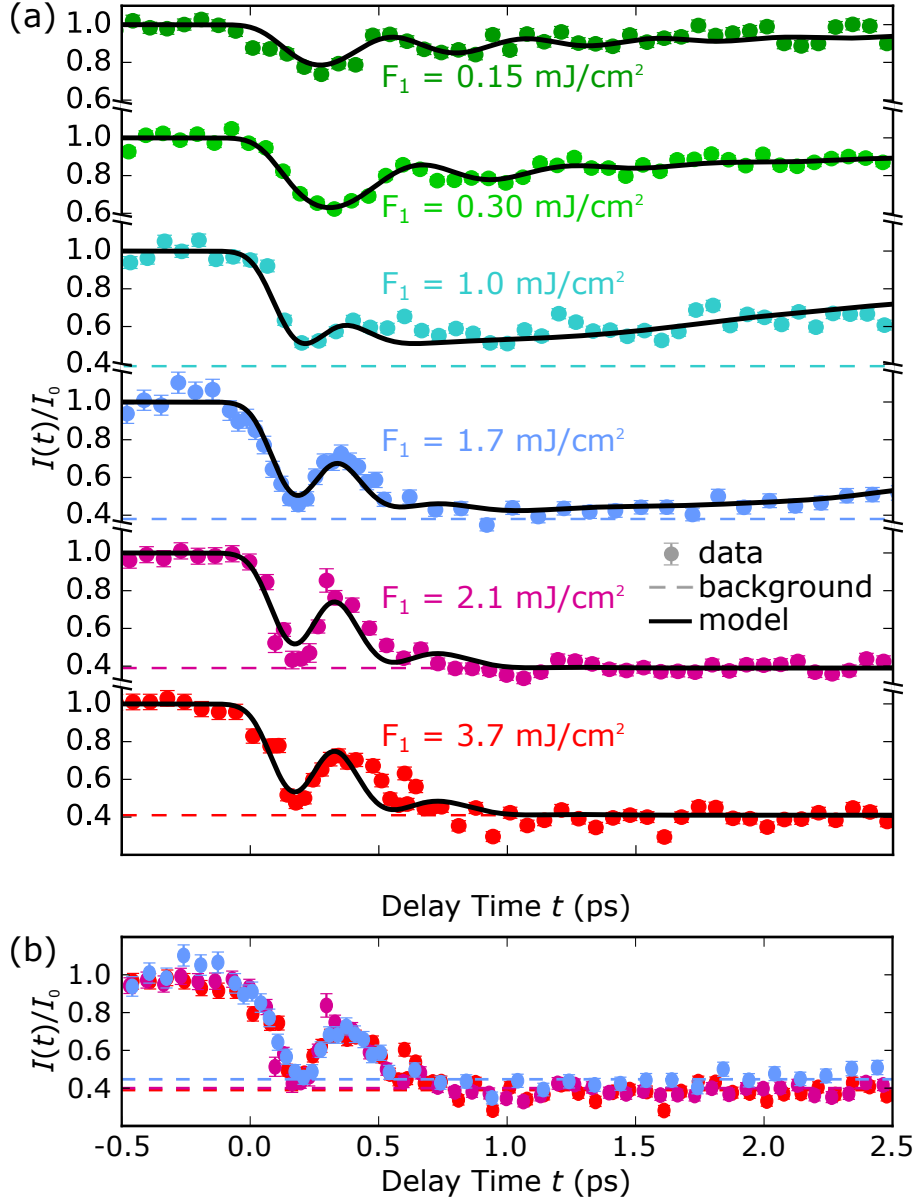}
\caption{\label{fig:F_scan} Single excitation (\(p_1\) only) with different fluences in (a). Solid curves are derived from the model presented in the main text. Dashed lines indicate the background levels. In (b) the data for \(F_1 = 1.7\), 2.1, and 3.7~mJ/cm\(^2\) are collapsed into one curve for comparison.}
\end{figure}

The top part of Fig.~\ref{fig:expl_plot}(a) shows a fit from this model compared to data with \(\Delta t\textsubscript{12}=0.30\)~ps and \(F_2 = F_1\), while the bottom part displays the time evolution of \(\gamma(t)\), and \(a(t)\) at \(z =0\). A sketch of the time-dependent potential energy surface is depicted in Fig. \ref{fig:expl_plot}(b). The letters A-E guide through the measured pump-probe dynamics relating the corresponding points in the potential landscape, while the background colors mark the current effective potential configuration. In the beginning the system is in its double-well equilibrium state at A. At \(t=0\) the first pump pulse \(p_1\) arrives, promotes \(a(t)\) to \(a\textsubscript{max}\) (B) and quenches \(\gamma(t)\) from \(\gamma\textsubscript{unex}\) to \(\gamma\textsubscript{min}\). The system then goes through the minimum and overshoots to the opposite side of the high-energy potential. At \(t=0.30\)~ps, \(p_2\) excites the system again (C), but does not change the shape of \(V\) and suppresses the damping \(\gamma(t)\). Afterwards, the system swings back to D, and finally comes to a stop in the single-well minimum at E, since the damping has in the meantime reached its maximal value \(\gamma_0\) (cf. bottom of Fig. \ref{fig:expl_plot}(a)).  

We fit all presented data sets with four global parameters, namely \(a\textsubscript{max}\), \(\gamma_0\), \(g\), and \(\tau_\gamma\), while \(\eta_0\) and \(\tau\textsubscript{disp}\) are determined only for the data sets showing a partial recovery within the monitored time frame (see Supplementary Information). \(I\textsubscript{bg}\) is fit for each curve individually (see Supplementary Information). The parameter \(\gamma_0 = 2.81\pm 0.38\)~ps\(^{-1}\) is similar to the damping constant close to the thermal transition (see Supplementary Information), whereas \(\tau_\gamma =0.18\pm 0.11\)~ps is comparable to the fast relaxation time of Ref. \cite{Tomeljak2009}. The resulting model curves are shown in all figures as black solid lines. With a small set of fit parameters our model reproduces the overall features of all data sets, including the single pump time traces at various fluences from Ref. \cite{Huber2014} (cf. Fig. \ref{fig:F_scan} (a)). The novel observation of the current double-pump data is that a second revival is present only when a second excitation arrives while the coherent motion after the first pump pulse still persists. This is well reproduced by our simple model, which furthermore describes the qualitative dynamics of the system quite consistently. This is true for both the absence of a second PLD revival for \(\Delta t\textsubscript{12}=1.00\)~ps and the scaling of \(A_2\) for different values of \(F_2\) as presented in Fig. \ref{fig:p_2_var}(b). 

The appearance of a second revival in the case of an additional pump between \(\Delta t\textsubscript{12}=0.18\) and 0.40~ps unambigously identifies this phenomenon as coherent PLD oscillations in the photoinduced high-symmetry phase. Starting from the model sufficient to explain superlattice dynamics triggered by a single-pulse excitation \cite{Huber2014}, we were able to refine the model and provide better understanding of the fundamental processes involved after exciting the electronic system. As mentioned above, the timing of the second revival is independent of changes in the timing and strength of the second pump pulse. This suggests that the frequency of the vibrational mode is not strongly changed by the second pulse. 

When comparing the presented PLD dynamics in the high symmetry phase to the doubly-pumped coherent structural dynamics of materials far from a phase transition, such as the coherently driven \(A_{1g}\) mode of bismuth at low excitation fluences \cite{Beaud2007, Hase1996}, a different behavior is noted. Here the symmetry of the potential energy surface is unchanged, allowing the second pulse to further shift the values of \(x\textsubscript{min}\) at well defined times \(t\) after the initial DECP.  This enables a selective enhancement or cancellation of the coherent phonon, since the effect of the second excitation depends on the phase of \(x(t)\). We observe something fundamentally different in the high excitation limit: The first pulse already changes the symmetry of the potential energy surface to that of the undistorted metallic phase, and the second pulse cannot further shift \(x\textsubscript{min}\) displacively. It does, however, influence the dynamics by extending the time over which underdamped dynamics occur. The mechanism behind the damping evolution is unclear, and could be either the result of a suppression of electron-phonon coupling channels or the modulation of anharmonic coupling to other vibrational modes. Methods like time- and angle-resolved photoelectron emission spectroscopy or non-equilibrium diffuse scattering could help to shed light on the details of the damping mechanism.
  
We have shown that we can sustain the coherent dynamics in the high-symmetry metallic phase of K\(_{0.3}\)MoO\(_3\) launched by strong electronic excitation with a femtosecond laser pulse through the phase transition, by re-exciting the system with additional pump pulses at slightly delayed times. We also note that this damping suppression is extremely efficient, as evidenced by the very large amplitude of the second PLD revivals seen in the experiment. Comparison with a simple phenomenological model suggests that the second excitation mainly manipulates the damping of the associated vibrational coordinate. While the exact mechanism remains unclear, the data is well fit using a damping whose magnitude depends on the delay between, and the strength of the excitation pulses. Thus, the coherence time of the reported oscillation can be extended by a second pump pulse. The fact that for optimal re-excitation conditions, at \(\Delta t\textsubscript{12} = 0.30\)~ps and \(F_2 = F_1\), the period and amplitude of the two resulting transient revivals are very similar indicates that the potential energy along the CDW distortion coordinate is largely unaffected by repeated excitation after crossing the transition to the metallic phase. We are therefore able to act upon the dynamics of the PLD associated with the CDW-phase even though the system has already undergone a photoinduced phase transition to its high-symmetry state.

\begin{acknowledgments}
Time resolved x-ray diffraction measurements were performed at the X05LA, and preparative static grazing incidence diffraction measurements were conducted at the X04SA beam lines of the Swiss Light Source, Paul Scherrer Institut, Villigen. We thank P. Willmott, and D. Grolimund for experimental help and L. Huber, and G. Lantz for discussion. We acknowledge funding through the NCCR Molecular Ultrafast Science and Technology (NCCR MUST), a research instrument of the Swiss National Science Foundation (SNSF). E. A. acknowledges support from the ETH Zurich Postdoctoral Fellowship and the Marie Curie Actions for People COFUND Programs, E. B.  from the European Community’s Seventh Framework Programme (FP7/2007-2013) under grant agreement No. 290605 (PSI-FELLOW/COFUND), and D. D. from the FemtoBias project, the Grant Agreement 55 of the NEWFELPRO fellowship project (Grant Agreement No. 291823) cofinanced by MSCA-FP7-PEOPLE-2011-COFUND.
\end{acknowledgments}

\pagebreak

\begin{LARGE}
\centering
\textbf{Supplementary Information}\\~\\
\end{LARGE}
\begin{Large}
Experimental Details\\
\end{Large}
At a grazing angle of 10\(^\circ\) the penetration depth of the 800~nm-pump is \(\delta_\textsubscript{L} = 80\)~nm, while for the x-rays at 0.4\(^\circ\) it is \(\delta\textsubscript{X} = 100\)~nm \cite{Huber2014}. The ultrashort x-ray pulses are generated by electron-beam slicing \cite{Beaud2007} and the intensity \(I\) of the \((1~(4-\nolinebreak[4]q\textsubscript{b})~\overline{0.5})\) Bragg peak is detected with an avalanche photodiode. The diameters of the spots of both pump beams on the sample are \(d\approx 500~\mu\)m, while the x-rays are focused vertically to 10~\(\mu\)m with a Kirkpatrick-Baez mirror and horizontally to 300~\(\mu\)m with a toroidal mirror \cite{Beaud2007}. The resulting temporal resolution is governed by the durations of the pump and probe pulses, their relative grazing angle and the extent of their respective spots on the sample.\\

\begin{Large}
Layer contributions\\
\end{Large}
To capture the inhomogeneous excitation profile of the 800~nm pump pulses, the probed volume with a depth of \(\delta\textsubscript{X}\) is split into ten layers with a thickness of \(d=10\)~nm each, like in Ref. \cite{Huber2014}. Like this the \(\eta_0\) of the \(j\)-th layer \(\eta_{0,j}\) is calculated as 
\begin{equation}
\eta_{0,j} = \eta_0 e^{-(j-1)d/\delta_\textsubscript{L}}.
\label{eq:eta_abs}
\tag{S1}
\end{equation}
The model x-ray intensity is then calculated as the weighed sum of the different layer intensity contributions with the weight \(\exp(-2jd/\delta\textsubscript{X})\) for the \(j\)-th layer.\\

\begin{Large}
Single Pump Fluence Dependence\\
\end{Large}
Fig. \ref{fig:F_scan}(a) shows the data from Ref. \cite{Huber2014} and one data set at \(F_1=1.7\)~mJ/cm\(^2\) from the current publication, with single excitation at different fluences. The displayed model curves are generated using the same methods described in the main text for Fig. 1-3. For the two curves with \(F_1 < 1.0\)~mJ/cm\(^2\) staying in the low symmetry configuration, the damping takes the form
\begin{equation}
\gamma(t) = 
\begin{cases}
\gamma\textsubscript{unex} & t < 0\\
\gamma_0 & t \geq 0,
\end{cases}
\label{eq:const_damp}
\tag{S2}
\end{equation}
with \(\gamma_0\) again being a fit parameter. The resulting model curves are shown as solid black lines and the respective background levels as dashed lines. For \(F_1\leq 1.0\)~mJ/cm\(^2\) the background level is set to 0.39.

The corresponding values of \(\eta_0\) for \(F_1 = 0.15\)~mJ/cm\(^2\), 0.30~mJ/cm\(^2\), and 1.0~mJ/cm\(^2\) are fit individually. To underline that the assumption of a limiting \(a\textsubscript{max}\) is well justified experimentally, Fig. \ref{fig:F_scan}(b) shows the three curves for \(F_1\geq 1.7\)~mJ/cm\(^2\) collapsed into one. Their similarity despite the fact that \(F_1\) is varied by more than a factor of two supports the assumption.\\

\begin{figure}
\includegraphics[width=8cm]{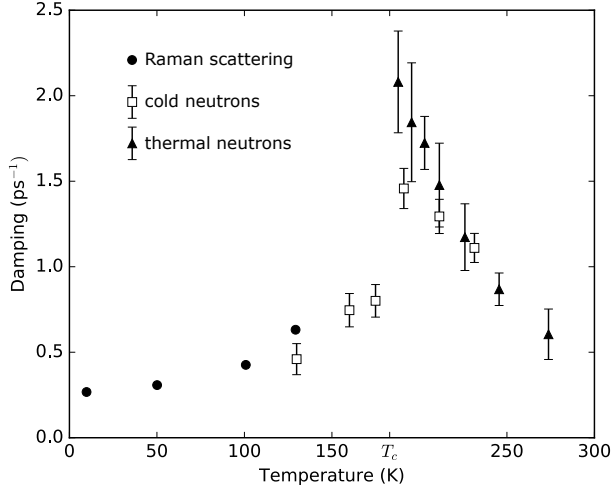}%
\caption{\label{fig:static_damp} Damping of the phonon mode associated with the CDW-formation in dependence of temperature measured with different experimental methods. The data is taken from Ref. \cite{Pouget1991}}
\end{figure}

\begin{Large}
Static damping\\
\end{Large}
Fig. \ref{fig:static_damp} shows the damping of the phonon mode that exhibits the Kohn anomaly and becomes the amplitude mode of the CDW below \(T\textsubscript{c}\). This damping was measured with different methods as marked in the figure, summarized in Ref. \cite{Pouget1991}. The value of 0.4~ps\(^{-1}\) for \(\gamma\textsubscript{unex}\) at 100~K is determined by these data, and the low limit \(\gamma\textsubscript{min}=0.2\)~ps\(^{-1}\) is based on an estimation for high temperature values.\\

\begin{Large}
Fit parameters\\
\end{Large}
The data sets presented in the main text and the supplement fall into different groups according to their features and experimental parameters. Table \ref{tab:groups} lists these groups together with the respective fit parameters and Table \ref{tab:sets} lists all data sets with their groups and parameters. Group I comprises all data sets. Within this group \(a\textsubscript{max}\) is held common. All data sets with \(F_1 = 1.7\)~mJ/cm\(^2\) that do show a partial recovery within the monitored time frame fall into group II and share the parameter \(\eta_0\). Notably, group II consists of all data sets with \(F_2 < F_1\), so 1.7~mJ/cm\(^2\) appears to be a threshold value for the onset of a recovery within 2.5~ps. All data sets that do show a partial recovery, i.e. also those with \(F_1 < 1.7\)~mJ/cm\(^2\), belong to group III, used to fit \(\tau\textsubscript{disp}\). Finally, group IV and V are those sets with and without the photoinduced phase transition. Like this, the damping parameters \(\gamma_0\), \(g\), and \(\tau_\gamma\) are shared within group IV, and so is \(\gamma_0\) within group V. 

\begin{table}[htb]
\centering
\begin{tabular}{|c|l|r|l|}
\hline
\multicolumn{1}{|c|}{group} & \multicolumn{1}{|c|}{par.} & \multicolumn{1}{|c|}{best fit} & \multicolumn{1}{|c|}{group description}\\
\hline
I & \(a\textsubscript{max}\) & 0.37 \(\pm\) 0.04 & all data sets\\
\hline
II & \(\eta_0\) & 2.11 \(\pm\) 0.25 & \(F_1 = 1.7\)~mJ/cm\(^2\), partial recovery visible\\
\hline
III & \(\tau\textsubscript{disp}\) (ps) & 3.08 \(\pm\) 0.67 & partial recovery visible\\
\hline
IV & 
	\begin{tabular}{@{}l@{}}\(\gamma_0\) (ps\(^{-1}\)) \\ \(g\) \\\(\tau_\gamma\) (ps)\end{tabular} &
	\begin{tabular}{@{}l@{}}2.81 \(\pm\) 0.38 \\ 4.39 \(\pm\) 0.54 \\ 0.18 \(\pm\) 0.11 \end{tabular} &
	phase transition\\
\hline
V & \(\gamma_0\) (ps\(^{-1}\))  & 1.24 \(\pm\) 0.51 & no phase transition\\
\hline
\end{tabular}
\caption{Groups of data sets according to different features relevant to the model curves including the respective fit parameters. The members of the groups can be found in Table \ref{tab:sets}}
\label{tab:groups}
\end{table}

\begin{table}
\centering
\begin{tabular}{|l|l|c|c|c|c|c|c|r|}
\hline
\multicolumn{1}{|c|}{Fig.} & \multicolumn{1}{|c|}{data set} & \multicolumn{1}{|c|}{\(a\textsubscript{max}\)} & \multicolumn{1}{|c|}{\(\eta_0\)} & \multicolumn{1}{|c|}{\(\tau\textsubscript{disp}\)} & \multicolumn{1}{|c|}{\(\gamma_0\)} & \multicolumn{1}{|c|}{g} & \multicolumn{1}{|c|}{\(\tau_\gamma\)} & \multicolumn{1}{|c|}{\(I\textsubscript{bg}\)}\\
\hline
1 (b) & \(p_1\) only & I & II & III & IV & IV & IV & 0.38 \(\pm\) 0.22\\
\hline
1 (b) & \(p_1\) and \(p_2\) & I & - & - & IV & IV & IV & 0.41 \(\pm\) 0.23\\
\hline
1 (b) & \(p_2\) only & I & II & III & IV & IV & IV & 0.39 \(\pm\) 0.19\\
\hline
2 (a) & \(\Delta t_{12} = 0.18\)~ps & I & - & - & IV & IV & IV & 0.30 \(\pm\) 0.19\\
\hline
2 (a) & \(\Delta t_{12} = 0.30\)~ps & I & - & - & IV & IV & IV & 0.38 \(\pm\) 0.19\\
\hline
2 (a) & \(\Delta t_{12} = 0.40\)~ps & I & - & - & IV & IV & IV & 0.33 \(\pm\) 0.19\\
\hline
2 (a) & \(\Delta t_{12} = 0.50\)~ps & I & - & - & IV & IV & IV & 0.34 \(\pm\) 0.19\\
\hline
2 (a) & \(\Delta t_{12} = 1.00\)~ps & I & - & - & IV & IV & IV & 0.29 \(\pm\) 0.16\\
\hline
2 (b) & \(F_2\) = 1/4 \(F_1\) & I & II & III & IV & IV & IV & 0.27 \(\pm\) 0.20\\
\hline
2 (b) & \(F_2\) = 1/2 \(F_1\) & I & II & III & IV & IV & IV & 0.32 \(\pm\) 0.20\\
\hline
2 (b) & \(F_2\) = \(F_1\) & I & - & - & IV & IV & IV & 0.30 \(\pm\) 0.20\\
\hline
2 (b) & \(F_1\) = 0.15~mJ/cm\(^2\) & I & \multicolumn{1}{|r|}{0.32 \(\pm\) 0.32} & III & V & - & - & \multicolumn{1}{|c|}{-}\\
\hline
2 (b) & \(F_1\) = 0.30~mJ/cm\(^2\) & I & \multicolumn{1}{|r|}{0.55 \(\pm\) 0.30} & III & V & - & - & \multicolumn{1}{|c|}{-}\\
\hline
2 (b) & \(F_1\) = 1.0~mJ/cm\(^2\) & I & \multicolumn{1}{|r|}{1.41 \(\pm\) 0.42} & III & IV & IV & IV & \multicolumn{1}{|c|}{-}\\
\hline
2 (b) & \(F_1\) = 2.1~mJ/cm\(^2\) & I & - & - & IV & IV & IV & 0.39 \(\pm\) 0.16\\
\hline
2 (b) & \(F_1\) = 3.7~mJ/cm\(^2\) & I & - & - & IV & IV & IV & 0.41 \(\pm\) 0.15\\
\hline
\end{tabular}
\caption{Table of all data sets presented with their respective model parameters. If a parameter is irrelevant for a certain data set the corresponding entry is "-", otherwise it either shows the best fit values including uncertainties or to which group of data sets the parameter is simultaneously fit. The best fit values for the parameters that are fit to more than one data set at a time are displayed in Table \ref{tab:groups}.}
\label{tab:sets}
\end{table}
\pagebreak
%

\end{document}